%% file: Dupertuis-LHCbTimeDependentProspects-CKMProceedings.tex
\documentclass[12pt]{article}
\usepackage{epsfig}
\usepackage{amssymb,amsmath}
\usepackage{float}

\input econfmacros.tex
\def\Title#1{\begin{center} {\Large {\bf #1} } \end{center}}

\def \BdJpsiKS{$B^0 \rightarrow J/\psi K^0_S$}
\def \BdJpsiKst{$B^0 \rightarrow J/\psi K^*$}
\def \BsJpsiKS{$B^0_s \rightarrow J/\psi K^0_S$}
\def \BsJpsiphi{$B^0_s \rightarrow J/\psi \phi$}
\def \BsJpsifzero{$B^0_s \rightarrow J/\psi f_0(980)$}
\def \BsJpsipipi{$B^0_s \rightarrow J/\psi \pi \pi$}
\def \Bsphiphi{$B^0_s \rightarrow \phi \phi$}
\def \BsJpsieta{$B^0_s \rightarrow J/\psi \eta$}
\def \BsJpsietap{$B^0_s \rightarrow J/\psi \eta'$}
\def \BsDsDs{$B^0_s \rightarrow D_s^+ D_s^-$}
\def \BsJpsiKst{$B^0_s \rightarrow J/\psi K^*$}
\def \BdJpsirho{$B^0 \rightarrow J/\psi \rho^0$}
\def \BdJpsiphi{$B^0 \rightarrow J/\psi \phi$}
\def \BdJpsipizero{$B^0 \rightarrow J/\psi \pi^0$}
\def \BdphiKS{$B^0 \rightarrow \phi K^0_S$}
\def \BdetapKS{$B^0 \rightarrow \eta' K^0_S$}

\def \BsDsK{$B^0_s \rightarrow D^{\mp}_s K^{\pm}$}
\def \BsDsKpipi{$B^0_s \rightarrow D^{\mp}_s K^{\pm} \pi^+ \pi^-$}
\def \BsKK{$B^0_s \rightarrow K^+ K^-$}
\def \BsKstKst{$B^0_s \rightarrow K^{*0} \bar{K}^{*0}$}
\def \Bsphifzero{$B^0_s \rightarrow \phi f_0(980)$}
\def \phis{$\phi_s$}
\def \phiseff{$\phi^{\mathrm{eff}}_s$}
\def \betas{$\beta_s$}

\def \sintwobeta{$\sin(2\beta)$}
\def \sintwobetaeff{$\sin(2\beta^{\mathrm{eff}})$}
\def \gammatwobetas{$\gamma - 2 \beta_s$}
\def \DGs{$\Delta \Gamma_s$}
\def \DGd{$\Delta \Gamma_d$}
\def \asls{$a^s_{\mathrm{sl}}$}
\def \asld{$a^d_{\mathrm{sl}}$}
\def \invfb{fb$^{-1}$}
\def \invpb{pb$^{-1}$}

\begin{document}

\Title{Prospects for time-dependent asymmetries at LHCb}

\bigskip\bigskip


\begin{raggedright}  

{\it Fr\'ed\'eric Dupertuis\index{Dupertuis, F.}\\
Laboratoire de Physique des Hautes Energies (LPHE), \\
Ecole Polytechnique F\'ed\'erale de Lausanne (EPFL),\\
CH-1015 Lausanne, SWITZERLAND. \\
On behalf of the LHCb collaboration.
}
\bigskip\bigskip
\end{raggedright}

\section{Introduction}

LHCb \cite{LHCbDetector} has been acquiring physics data since 2010 and recorded about 0.04~\invfb\ in 2010 and 1.1~\invfb\ in 2011 at center-of-mass energy ($\sqrt{s}$) of 7 TeV. In 2012, it is projected to record 2.2~\invfb\ at an energy of $\sqrt{s} = 8$ TeV with a nominal instantaneous luminosity of $\mathcal{L} = 4\times10^{32}$ cm$^{-2}\,$s$^{-1}$, before a long shutdown of almost two years. The data taking is expected to resume by the end of 2014 at $\sqrt{s}=13-14$ TeV, before a second long shutdown in 2018 when the upgraded LHCb detector components will be installed. Since the $b\bar{b}$ cross section depends almost linearly on $\sqrt{s}$, this will lead to an increase of about 100\% in $b\bar{b}$ pairs yield at $\sqrt{s} = 14$ compared to $\sqrt{s} = 7$. By 2018, a data sample larger than 8~\invfb\ is expected to have been recorded, leading to an increase of about a factor four in statistical power with respect to the 1~\invfb\ sample recorded at $\sqrt{s} = 7$ TeV.

\begin{figure}[htb]
\begin{center}
\includegraphics[width=0.35\columnwidth,keepaspectratio]{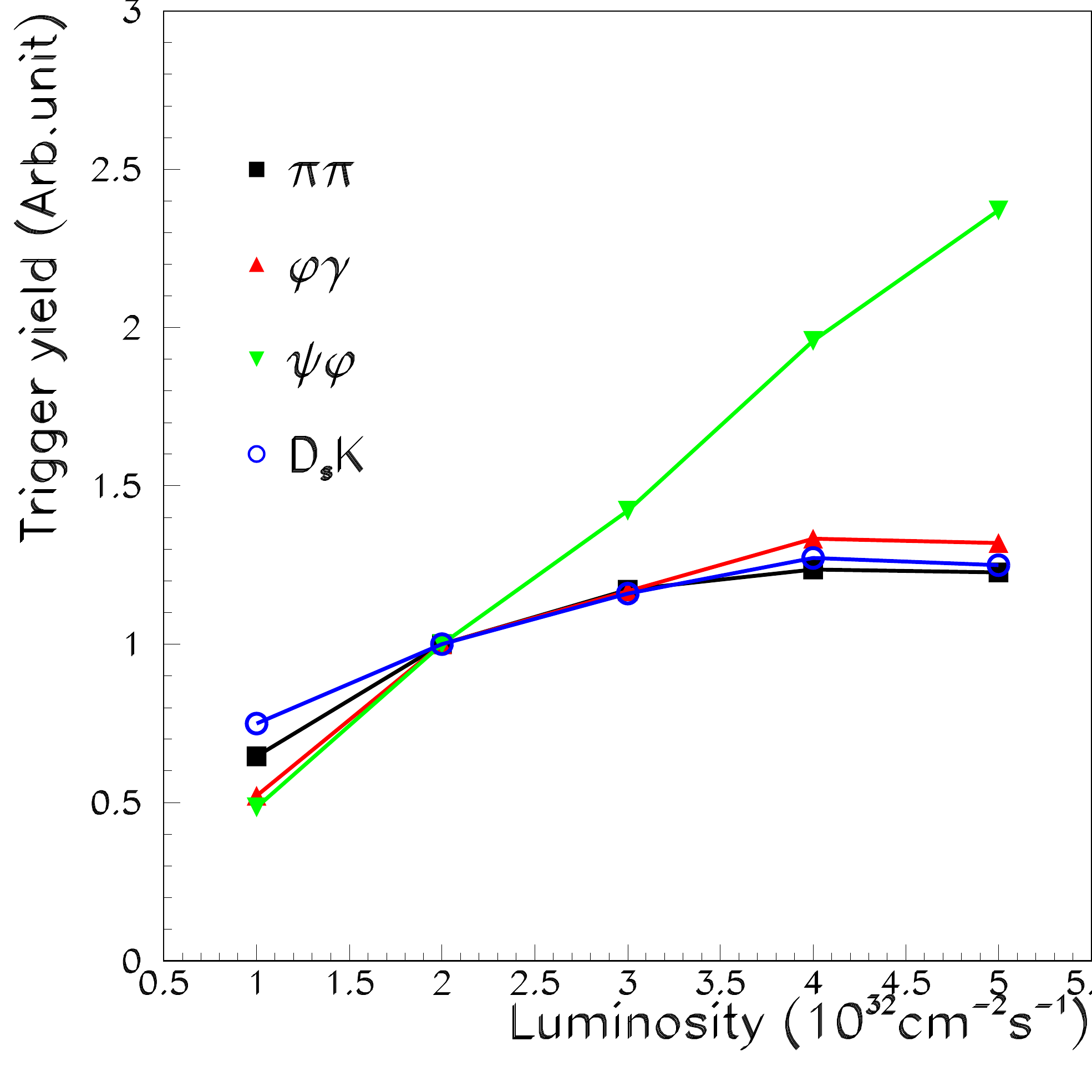}
\hspace{1cm}
\includegraphics[width=0.53\columnwidth,keepaspectratio]{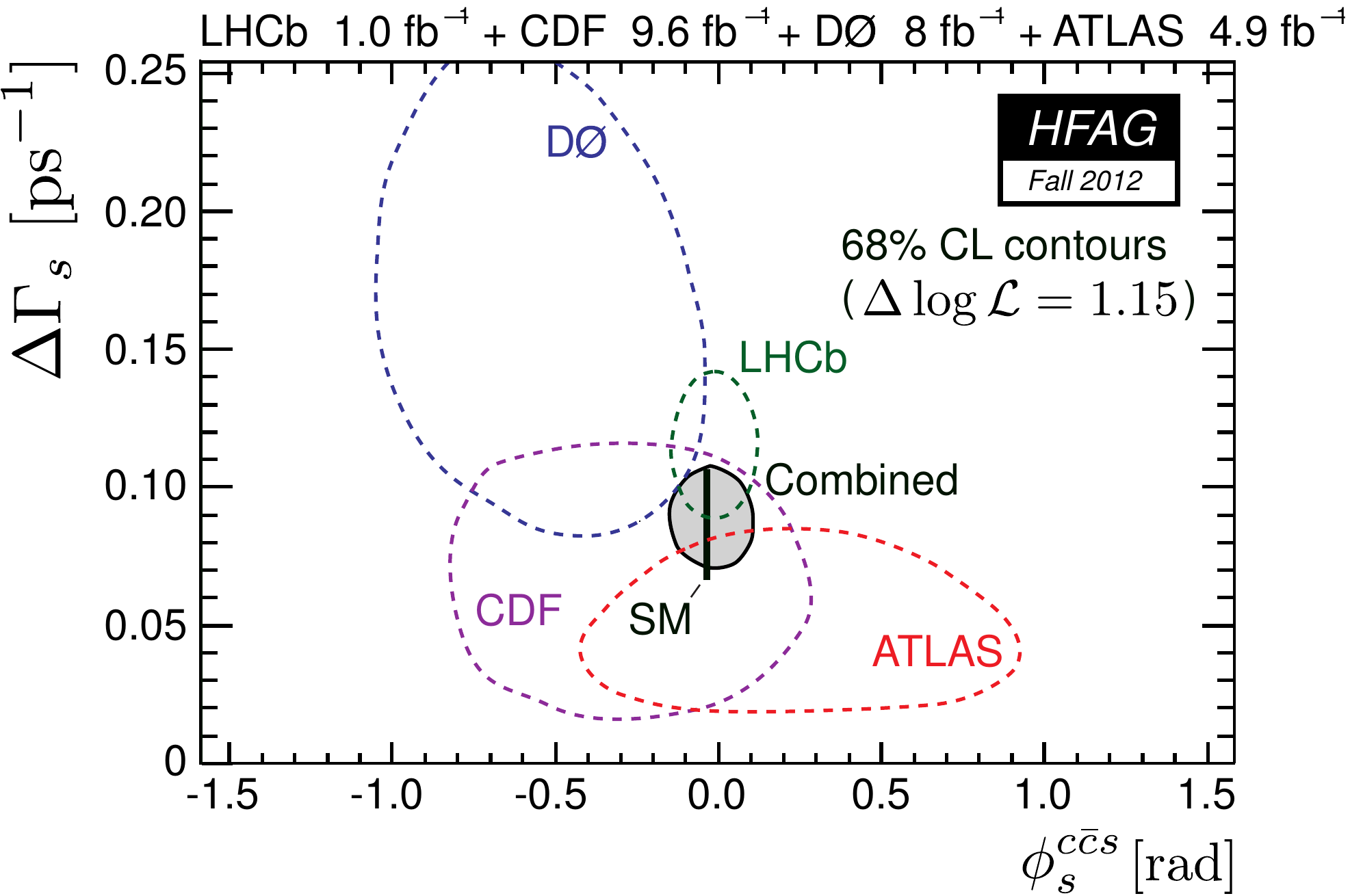}
\caption{\textit{Left}: Trigger yield as function of the instantaneous luminosity for different decay channels with the current LHCb trigger system. The trigger yield for hadronic modes ($\pi\pi,\phi\gamma,D_sK$) flatens compared to modes with muons in their final states ($\psi\phi$). \textit{Right}: Global fit of $\Delta\Gamma_s-\phi_s$ results performed by HFAG including the preliminary results of LHCb with 1~\invfb\ of data.}
\label{fig1}
\end{center}
\end{figure}

The LHCb upgrade \cite{LHCbUpgradeLoI} is designed to take data up to a luminosity of $\mathcal{L} = 2 \cdot 10^{33}\, \mathrm{cm}^{-2}\,\mathrm{s}^{-1}$ at $\sqrt{s} = 13-14$ TeV, recording more than 5 \invfb\ each year. In order not to suffer from large pile-up, the 25~ns bunch spacing of the LHC will be required. The detector readout will be upgraded to allow the full 40 MHz LHC interaction rate to be read into a software trigger, improving the trigger efficiency on hadronic modes a factor 2 (Figure \ref{fig1}). With an operation time of 10 years starting from 2019, it is expected to record more than 50~\invfb. That will lead to a gain in statistical power by a factor ten with respect to the 1~\invfb\ of LHCb. This paper are based on the LHCb upgrade Letter-Of-Intent (LoI) \cite{LHCbUpgradeLoI}, the LHCb upgrade Framework TDR \cite{FTDR} and the LHCb prospects paper \cite{LHCbProspects}. A summary of the status and the prospects of the time-dependent $CP$-observables is provided with their statistical error expectations. Systematic errors are expected to be kept below the statistical ones throughout the upgrade programme.

\section{Weak phase in tree-dominant $b\rightarrow c \bar{c}s$ transitions}

$B^0/B^0_s$-mesons decaying to $CP$-eigenstates through tree-dominant $b\rightarrow c \bar{c}s$ transitions give rise to a weak phase, coming from the interference between the direct decay and the decay after mixing. For \BsJpsiphi\ and \BsJpsipipi\ decays the weak phase \phis\ is equal to $\phi_{s,M} - 2\, \phi_{s,D}$ where $\phi_{s,M}$, $\phi_{s,D}$ are the weak phases rising in the mixing and decay, respectively. Within the SM, $\phi_{s,M}^{\mathrm{SM}} = - 2 \beta_s$ and $\phi_{s,D}^{\mathrm{SM}} = \phi_{s}^{\mathrm{SM,penguin}}$ where \betas\ is the weak phase associated to the CKM element $V_{ts}$, which is predictaed to be small $-2 \beta_s = (-0.036 \pm 0.002)\ \mathrm{rad}$ \cite{LHCbProspects}, and $\phi_{s}^{\mathrm{SM,penguin}}$  is the phase coming from the doubly Cabibbo suppressed penguin decay involving $V_{ub}$ which is expected to be even smaller than \betas. In the presence of New Physics (NP), the measured weak phase $\phi_s$ can be significantly enlarged with respect to the very small contribution from the SM. The full tagged angular time-dependent analysis of \BsJpsiphi\ and tagged time-dependent analysis of \BsJpsipipi\ allow a precision measurement of $\phi_s$. With 1~\invfb\ of data, results of \phis\ from a preliminary analysis of \BsJpsiphi\ decays and from \BsJpsipipi\ decays give: 
\begin{center}
 $\phi^{J/\psi\phi}_s = -0.001 \pm 0.101\, (\mathrm{stat}) \pm 0.027\, (\mathrm{syst})\, \mathrm{rad}$ \cite{BsJpsiphiMoriond}, \\
 $\phi^{J/\psi\pi\pi}_s = -0.019\, ^{+0.173}_{-0.174}\, (\mathrm{stat})\, ^{+0.004}_{-0.003}\, (\mathrm{syst})\, \mathrm{rad}$ \cite{BsJpsipipi}.
\end{center}

The combined result is $\phi_s = -0.002 \pm 0.083\, (\mathrm{stat}) \pm 0.027\, (\mathrm{syst})\, \mathrm{rad}$ \cite{BsJpsiphiMoriond} (Figure \ref{fig1}). These results are compatible with the SM prediction and more data is required to pin down possible small NP effects. 

The expected precisions for measuring $\phi_s$ are 0.025 (0.045) rad with the full LHCb dataset and 0.008 (0.014) rad with LHCb upgrade where the theoretical uncertainty of 0.003 (0.01) rad for \BsJpsiphi\ (\BsJpsipipi) will allow sensitivity to small NP phenomena. These results will be compared to measurements of \phis\ in different decay modes such as \BsJpsieta, \BsJpsietap\ and \BsDsDs\ which have more significant penguin contributions. With 1~\invfb, preliminary results yields $477 \pm 23$ \BsDsDs\ candidates ($\mathcal{B}(\bar{B}^0_s\rightarrow D^+_s D^-_s)/\mathcal{B}(\bar{B}^0\rightarrow D^+ D^-_s) = 0.508 \pm 0.026\, (\mathrm{stat}) \pm 0.043\, (\mathrm{syst})$) \cite{BsDsDs} so the \phis\ measurement in these decays will soon commence at LHCb. A clean measurement of \phis\ can be obtained by performing a tagged time-dependent Dalitz-plot analysis of $B^0_s \rightarrow \bar{D}^0(\rightarrow KK)KK$ penguin-free decays. With 0.62~\invfb, $104 \pm 29$ $B^0_s \rightarrow \bar{D}^0(\rightarrow K\pi)KK$ candidates ($\mathcal{B} = (4.7 \pm 0.9\, (\mathrm{stat}) \pm 0.6\, (\mathrm{syst}) \pm 0.5\, (f_s/f_d)) \cdot 10^{-5}$) \cite{BsDKK} have been found which would yield to about 10 $B^0_s \rightarrow \bar{D}^0(\rightarrow KK)KK$. A measurement of $\phi_s$ in this channel can be attempted with the LHCb upgrade.

In the $B^0$ sector, a tagged time-dependent analysis of \BdJpsiKS\ decays allows a measurement of the weak phase $\phi$ that gives within the SM almost the CKM angle $\beta$. As for \phis, a small pollution $\phi^{\mathrm{SM,penguin}}$ from penguin diagrams is expected. New NP particles entering the loop in penguin diagrams can give rise to new phase $\phi^{\mathrm{NP}}$ such that the measured $\phi = 2\beta + \phi^{\mathrm{SM,penguin}} + \phi^{\mathrm{NP}}$. With 1~\invfb, the result of \sintwobeta\ from the LHCb analysis of \BdJpsiKS\ decays was presented at the CKM workshop \cite{sintwobetaCKM}, $\sin(2\beta) = 0.73 \pm 0.07\, (\mathrm{stat}) \pm 0.04\, (\mathrm{syst})$ \cite{sintwobeta}. By the end of 2012, LHCb will be able to provide competitive results on the single measurement of \sintwobeta\ with $B^0 \rightarrow J/\psi (\rightarrow \mu^+ \mu^-) K^0_S (\rightarrow \pi^+ \pi^-)$ and with the full LHCb dataset, the world's best measurement is expected. With the LHCb upgrade the statistical error on \sintwobeta\ is expected to go down to 0.0035. A clean measurement of \sintwobeta\ can be obtained by performing a full tagged time-dependent Dalitz-plot analysis of $B^0_s \rightarrow \bar{D}^0(\rightarrow KK)\pi\pi$ penguin-free decays. With 0.62~\invfb, $8060 \pm 29$ $B^0_s \rightarrow \bar{D}^0(\rightarrow K\pi)\pi\pi$ candidates \cite{BsDKK} have been found which would translate to about 800 $B^0_s \rightarrow \bar{D}^0(\rightarrow KK)\pi\pi$.

The penguin pollution in $b\rightarrow c \bar{c}s$ transitions is expected to be smaller than the weak phase arising from the tree processes within the SM. But with the precision that LHCb is aiming at for \phis\ and \sintwobeta, and furthermore with the LHCb upgrade, it becomes mandatory to have a handle on the weak phase coming from the penguin process. This can be assessed by performing the analysis of SU(3)-related decays such as \BsJpsiKst, \BdJpsirho\ or \BdJpsiphi\ in the case of \BsJpsiphi, and \BsJpsiKS\ or \BdJpsipizero\ channels in the case of \BdJpsiKS. With 0.37~\invfb, $116 \pm 13$ \BsJpsiKS\ candidates ($\mathcal{B} = (1.83 \pm 0.21\, (\mathrm{stat}) \pm 0.10\, (\mathrm{syst}) \pm 0.14\, (f_s/f_d) \pm 0.07\, (\mathcal{B}(B^0\rightarrow J/\psi K^0))) \cdot 10^{-5}$) \cite{BsJpsiKS} have been observed and with 0.41~\invfb, $114 \pm 11$ \BsJpsiKst\ candidates ($\mathcal{B} = (4.4 \,^{+0.5}_{-0.4}\, (\mathrm{stat}) \pm 0.8\, (\mathrm{syst})) \cdot 10^{-5}$) \cite{BsJpsiKst} have been observed. This means that with complete analysis of these channels, LHCb will be able to provide first constraints on the penguin pollution size with the dataset recorded up to the end of 2012.

\section{Weak phase in penguin-only $b\rightarrow s q\bar{q}$ with $q = s,d$ transitions}

In the $B^0_s$ sector, $b\rightarrow s q\bar{q}$ transitions with $q = s,d$ are penguin-only processes. With the dominant $t$-loop there is a cancellation between the weak phase in the mixing and in the decay, so the measured weak phase \phiseff\ vanishes within the SM. \phiseff\ can be extracted from a tagged angular time-dependent analysis of \Bsphiphi\ and \BsKstKst\ decays and a tagged time-dependent analysis of \Bsphifzero. 

An untagged angular time-integrated analysis of \Bsphiphi\ candidates has been performed to measure the triple product asymmetries $A_U, A_V$ where $801 \pm 29$ \Bsphiphi\ have been observed with a negligible $S$-wave fraction of $(1.3 \pm 1.2)\%$ \cite{BsPhiPhi}. Compatible with a vanishing $\phi_s^{\mathrm{eff}}$ they have been measured with:
\begin{center}
 $A_U = -0.055 \pm 0.036\, (\mathrm{stat}) \pm 0.018\, (\mathrm{syst})$ \cite{BsPhiPhi},\\ 
 $A_V = +0.010 \pm 0.036\, (\mathrm{stat}) \pm 0.018\, (\mathrm{syst})$ \cite{BsPhiPhi}.
\end{center}

With 36 \invpb, $50 \pm 7$ \BsKstKst\ candidates ($\mathcal{B} = (2.81 \pm 0.46\, (\mathrm{stat}) \pm 0.45\, (\mathrm{syst}) \pm 0.34\, (f_s/f_d)) \cdot 10^{-5}$) have been found with a $CP$-averaged $K^{*0}$ longitudinal polarisation fraction of $f_L = 0.31 \pm 0.12\, (\mathrm{stat}) \pm 0.04\, (\mathrm{syst})$ \cite{BsKstKst}. A tagged angular time-dependent analyses of these decays are expected with the 2011-2012 dataset and the precision on $\phi_s^{\mathrm{eff}}$ with 1~\invfb\ has been estimated to be around 0.3-0.4 rad with the channel \Bsphiphi\ and 0.03 rad from 50~\invfb\ of data with the LHCb upgrade experiment.

In the $B^0$ sector, the weak phase \sintwobetaeff\ rising from $b\rightarrow s q\bar{q}$ processes with $q = s,d$ is not far from \sintwobeta\ within the SM, with $|\Delta S| = |\sin(2\beta^{\mathrm{eff}}) - \sin(2\beta)| \lesssim 0.1$ \cite{LHCbProspects}. The current single best measurements of \sintwobetaeff\ come from the $B$-factories with a precision of about 0.08-0.10 using \BdetapKS\ decays. LHCb is currently studying to the \BdphiKS\ decays and the precision on \sintwobetaeff\ is expected to go down to 0.06 with 50~\invfb.

\section{Weak phase \gammatwobetas\ with \BsDsK}

Performing a tagged time-dependent analysis of \BsDsK\ decays allows the weak phase \gammatwobetas\ to be measured, where $\gamma = \arg[-V_{ud}V^*_{ub}/(V_{cd}V^*_{cb})]$. First preliminary results of a fit to the $CP$-observables of \BsDsK\ with 1~\invfb\ have been presented at the CKM workshop \cite{BsDsKCKM,BsDsK}. The precision on $\gamma$ from \BsDsK\ is expected to be about 11$^\circ$ with the full LHCb dataset and 2$^\circ$ with the LHCb upgrade. \BsDsKpipi\ decays are also expected to provide constraints on $\gamma$.

\section{Decay width difference \DGs\ and \DGd}

The best measurement of the decay width difference \DGs\ comes from the preliminary results of the tagged angular time-dependent analysis of \BsJpsiphi\ decays with 1~\invfb\ which gives $\Delta \Gamma_s = 0.116 \pm 0.018\, (\mathrm{stat}) \pm 0.006\, (\mathrm{syst})\, \mathrm{ps}^{-1}$ \cite{BsJpsiphiMoriond}. The error on \DGs\ is expected to go down to 0.003 ps$^{-1}$ with 50~\invfb. It is possible to constrain $\Delta \Gamma_s - \phi_s$ by combining the effective lifetimes of \BsJpsifzero\ ($CP$-even) and \BsKK\ ($CP$-odd). The sign of \DGs\ has been resolved at 4.7$\sigma$ to be positive using the running of the phase difference between the $P$-wave and $S$-wave around the $\phi(1020)$ resonance \cite{SignDeltaGammas}.

LHCb expects to measure \DGd\ with a precision of 0.02 ps$^{-1}$ with 1~\invfb\ by comparing the effective lifetimes of \BdJpsiKS\ and \BdJpsiKst, and 0.002 ps$^{-1}$ with 50~\invfb.

\section{Semileptonic asymmetries \asls\ and \asld}

\begin{minipage}{0.55\textwidth}
 The flavour-specific asymmetry \asls\ can be measured by computing the asymmetry between $B^0_s \rightarrow D^+_s \mu^- X$ and $B^0_s \rightarrow D^-_s \mu^+ X$ yields with $D_s^\pm \rightarrow \phi (\rightarrow K^+ K^-) \pi^\pm$ (use of other $D_s$ decays is foreseen). \asld\ can be similarly measured using $B^0 \rightarrow D^\pm \mu^\mp X$ decays with $D^\pm \rightarrow K^\mp \pi^\pm \pi^\pm$. The first LHCb results of \asls\ with 1~\invfb\ gives:
 \begin{center}
  $a^s_\mathrm{sl} = (-0.24 \pm 0.54\, (\mathrm{stat}) \pm 0.33\, (\mathrm{syst}))\%$ \cite{asls}.
 \end{center}
With the full LHCb dataset, a statistical precision of about 0.06\% is expected and 0.02\% with 50~\invfb.
\end{minipage}
\hspace{0.4cm}
\begin{minipage}{0.4\textwidth}
\begin{figure}[H]
\begin{center}
\includegraphics[width=0.8\columnwidth,keepaspectratio]{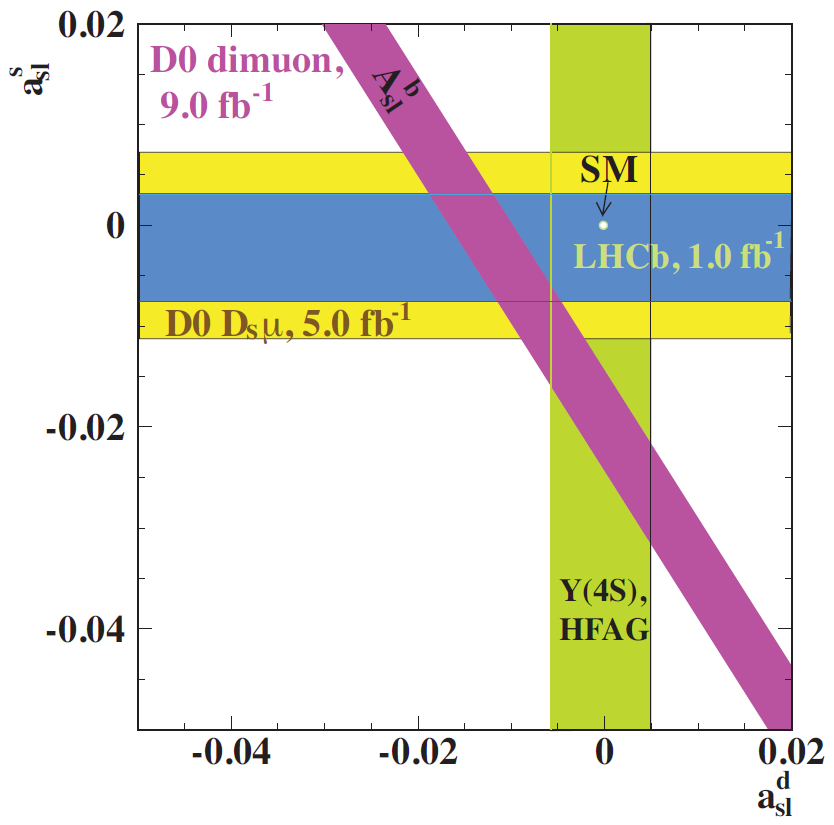}
\caption{$a^s_{\mathrm{sl}}-a^d_{\mathrm{sl}}$ results including the LHCb results with 1~\invfb\ \cite{asls}.}
\label{fig2}
\end{center}
\end{figure}
\end{minipage}

\section{Conclusion}

Table \ref{expectations} summarises the current precision and the expectations of statistical accuracy of the $CP$ observables with the LHCb detector and the LHCb upgrade. LHCb results are already providing world's best measurements in the $B^0_s$ sector. The LHCb upgrade will provide high precision measurements that will stringently test the CKM paradigm and contribute significantly to the search for possible small NP effects.

\begin{table}[htb]
\begin{center}
\begin{tabular}{|c|c|c|c|c|}
 \hline
  Parameter & LHCb 2011 & LHCb 2018 & Upgrade & Theory \\
  & (1~\invfb) & (8~\invfb) & (50~fb$^{-1})$ & uncertainty \\
  \hline
 $\phi_s$(\BsJpsiphi) [rad] & 0.10 & 0.025 & 0.008 & 0.003 \\
 $\phi_s$(\BsJpsipipi) [rad] & 0.17 & 0.045 & 0.014 & 0.01 \\
 \hline
 $\sin(2\beta)$(\BdJpsiKS) [-] & 0.07 & - & 0.0035 & 0.02 \\
 \hline
 $\phi^{\mathrm{eff}}_s$(\Bsphiphi) [rad] & \textit{0.3-0.4 (est.)} & - & 0.03 & $<$ 0.02 \\
 \hline
 $\sin(2\beta^{\mathrm{eff}})$(\BdphiKS) [-] & - & - & 0.06 & 0.02 \\
 \hline
 $\gamma$(\BsDsK) (t) [$^\circ$] & - & 11$^\circ$ & 2$^{\circ}$ & negligible \\
 \hline
 $\Delta \Gamma_s$(\BsJpsiphi) [ps$^{-1}$] & 0.018 & - & 0.003 & 0.02 \\
 \hline
 $\Delta \Gamma_d$($B^0 \rightarrow J/\psi K^0_S/K^*$) [ps$^{-1}$] & \textit{0.02 (est.)} & - & 0.002 & 0.001 \\
 \hline
 $a^s_{\mathrm{sl}}$($B^0_s \rightarrow D^\pm_s \mu^\mp X$) [\%]  & 0.54 & 0.06 & 0.02 & 0.003 \\
 \hline
\end{tabular}
\caption{Current and expected statistical accuracy of the relevent $CP$-observables measurable with time-dependent analyses with LHCb and the LHCb upgrade. The current theory uncertainties are also provided. \cite{FTDR,LHCbProspects}}
\label{expectations}
\end{center}
\end{table}



\end{document}

%% file: econfmacros.tex
\def\beq{\begin{equation}}
\def\eeq#1{\label{#1}\end{equation}}
\def\eeqn{\end{equation}}

\def\beqa{\begin{eqnarray}}
\def\eeqa#1{\label{#1}\end{eqnarray}}
\def\eeqan{\end{eqnarray}}

\let\bar=\overbar

\def\Dslash{\not{\hbox{\kern-4pt $D$}}}
\def\dslash{\not{\hbox{\kern-2pt $\del$}}}

\def\msb{{\bar{\ssstyle M \kern -1pt S}}}

